\documentclass[twocolumn]{aastex631}

\usepackage{xcolor}
\usepackage{xspace}

\usepackage{xcolor}

\newcommand{\gaia}{\emph{Gaia}\xspace}
\newcommand{\hst}{\emph{HST}\xspace}
\newcommand{\hip}{\emph{Hipparcos}\xspace}
\newcommand{\onefour}{HSToC-1\xspace}
\newcommand{\twoone}{HSToC-2\xspace}
\newcommand{\twothree}{HSToC-3\xspace}
\newcommand{\twonine}{HSToC-4\xspace}
\newcommand{\threeone}{HSToC-5\xspace}

\begin{document}
 
\title{Discovery of astrometric accelerations by dark companions in the globular cluster $\omega$ Centauri}

\correspondingauthor{Imants Platais, Johannes Sahlmann}
\email{imants@jhu.edu, Johannes.Sahlmann@ext.esa.int}

\author[0000-0003-2599-2459]{Imants Platais}
\affiliation{Department of Physics and Astronomy, Johns Hopkins University, 3400 North Charles Street, Baltimore, MD 21218, USA}

\author[0000-0001-9525-3673]{Johannes Sahlmann}
\altaffiliation{This author shares the lead authorship}
\affiliation{RHEA Group for the European Space Agency (ESA), European Space Astronomy Centre (ESAC),\\ Camino Bajo del Castillo s/n, 28692 Villanueva de la Ca\~nada, Madrid, Spain}

\author[0000-0002-6301-3269]{L\'eo Girardi}
\affiliation{Osservatorio Astronomico di Padova - INAF, Vicolo dell'Osservatorio 5, I-35122 Padova,Italy}

\author[0000-0003-0218-386X]{Vera Kozhurina-Platais}
\affiliation{Space Telescope Science Institute, 3700 San Martin Drive, Baltimore, MD 21218, USA}

\author[0000-0001-6604-0505]{Sebastian Kamann}
\affiliation{Astrophysics Research Institute, Liverpool John Moores University, IC2 Liverpool Science Park, 146 Brownlow Hill, Liverpool L3 5RF, UK}

\author{Dimitri Pourbaix}
\altaffiliation{Deceased on 14 November 2021}
\affiliation{FNRS, Institut d’Astronomie et d’Astrophysique, Université Libre de Bruxelles, Boulevard du Triomphe, 1050 Bruxelles, Belgium}

\author[0009-0006-5696-7706]{Florence Wragg}
\affiliation{Astrophysics Research Institute, Liverpool John Moores University, IC2 Liverpool Science Park, 146 Brownlow Hill, Liverpool L3 5RF, UK}

\author{Gerard Lemson}
\affiliation{Department of Physics and Astronomy, Johns Hopkins University, 3400 North Charles Street, Baltimore, MD 21218, USA}

\author{Arik Mitschang}
\affiliation{Department of Physics and Astronomy, Johns Hopkins University, 3400 North Charles Street, Baltimore, MD 21218, USA}

\begin{abstract}
We present results from the search for astrometric accelerations of stars in $\omega$ Centauri using 13 years of regularly-scheduled {\it Hubble Space Telescope} WFC3/UVIS calibration observations in the cluster core. The high-precision astrometry of $\sim$160\,000 sources was searched for significant deviations from linear proper motion. This led to the discovery of four cluster members and one foreground field star with compelling acceleration patterns. We interpret them as the result of the gravitational pull by an invisible companion and determined preliminary Keplerian orbit parameters, including the companion's mass. {For the cluster members} our analysis suggests periods ranging from 8.8 to 19+ years and dark companions in the mass range of $\sim$0.7 to $\sim$1.4$M_{\sun}$. At least one companion could exceed the upper mass-boundary of white dwarfs and can be classified as a neutron-star candidate.
\end{abstract}


\section{Introduction}\label{sec:intro}
Since the beginning of measuring proper motions it is implied that individual stars move along straight lines. This is still a good approximation for most sources even now in the \gaia era. If the motion indicates curvilinear pattern, that would change the proper motion by astrometric acceleration, usually measured in mas~yr$^{-2}$.  Stronger accelerations may allow to invoke the Keplerian formalism \citep[e.g.][]{jia19}. Thus, the detection of astrometric accelerations near the radio source Sagittarius A$^{*}$ was instrumental to eliminate all other alternatives to the supermassive black hole \citep{ghe00,2018A&A...615L..15G}. This discovery and additional astrometric accelerating sources in the same area {are examples} of ground-based astrometry {of binary motion where one component is dark}.

{In 1990-1993 the \hip space astrometry mission regularly observed  multiple times $\sim$118,000 pre-selected stars. For the first time in history regular astrometric observations were obtained outside the Earth's atmosphere and covering the entire celestial sphere. This effort resulted in 2622 acceleration candidates \citep{1997A&A...323L..53L} but none with a genuine invisible companion.}

The third data release (DR3) of the \gaia space astrometry mission used data acquired between July 2014 and May 2017 and reported 338,215 acceleration solutions and 165,500 orbital solutions \citep{hal22}. A few illustrative cases of compact and invisible companions are presented by the \gaia consortium in \citet{are22}. To `weigh` an
invisible companion one must determine the mass of the primary source and it is desirable to obtain epoch radial velocities. This recipe made it possible to discover Gaia BH1 and Gaia BH2, two dormant black holes (BH) in intermediate-period binaries \citep{elb23a, elb23b}.  
An opportunity to enhance the accuracy of proper motions and to probe very long-period accelerations is the combination of the \hip and \gaia catalogs \citep[e.g.][]{bra21}.  

Although the {\it Hubble Space Telescope} (\hst) is not an astrometric
observatory, it has provided spectacular science results such as the detection
of the very small tangential velocity of the Andromeda galaxy at 17.0 km s$^{-1}$
\citep{van12}, establishing the upper limit of a putative,
intermediate-mass black hole in $\omega$~Cen \citep{van10}, precision parallaxes ($\pm$45~$\mu$as) for long-period Cepheids in the Milky Way with the spatial scanning technique \citep{2018ApJ...855..136R}, and the detection of astrometric microlensing by a black hole \citep{2022ApJ...933...83S}.

These breakthroughs gave us confidence that measurements of 
astrometric accelerations are feasible with the prime \hst WFC3/UVIS camera. One configuration that can generate a measurable acceleration in a few years is a double star with a faint or dark secondary component.  
In order to maximize the number of surveyed sources and therefore the odds of detection, we focused on dense stellar system such as the core of globular clusters. Thanks to the foresight of the \hst WFC3 team, a multi-cycle calibration program targeting the center of $\omega$~Cen (NGC~5139) was set in motion in 2010. So far, a total of $\sim$200 well time-spread frames have been accumulated, spanning over a total of 13 years. In certain
aspects these observations may surpass the available \gaia catalogues due to the
2-dimensional pointed observations, the longer time span, and the slightly better spatial resolution which is critical in very crowded stellar environments such as the cores of globular clusters. However, the \emph{Gaia} limitations due to source blending and onboard resources discussed in \cite[][Sect. 6.6]{GaiaCollaboration:2016aa} were partially mitigated by a special observing mode \citep[e.g.][]{Sahlmann:2016ac} and a focused product release in the $\omega$~Cen region {was published on 10 October 2023 \citep{2023arXiv231006551G}}, see Sect.\ \ref{sec:fpr}.

A similar \hst program (GO-12911, PI L.R. Bedin) targeted the nearest ($\sim$1.7~kpc) Galactic globular cluster M~4 \citep[NGC~6121][]{bed13}. However, the timespan of these observations is only one year with an expectation that this would be sufficient to detect the astrometric `wobble`.  As of this writing, no accelerating source in M~4 has been
reported. The combined analysis of proper motions from \hst and \gaia DR3 claim a central dark mass of 800 $M_\sun$ in M~4 \citep{vit23}. 

\section{\hst/WFC3 OBSERVATIONS OF $\omega$ Cen}\label{hstobs}
\begin{figure}[ht!]
\includegraphics[trim={5mm 5mm 5mm 5mm}, clip, width=\columnwidth]{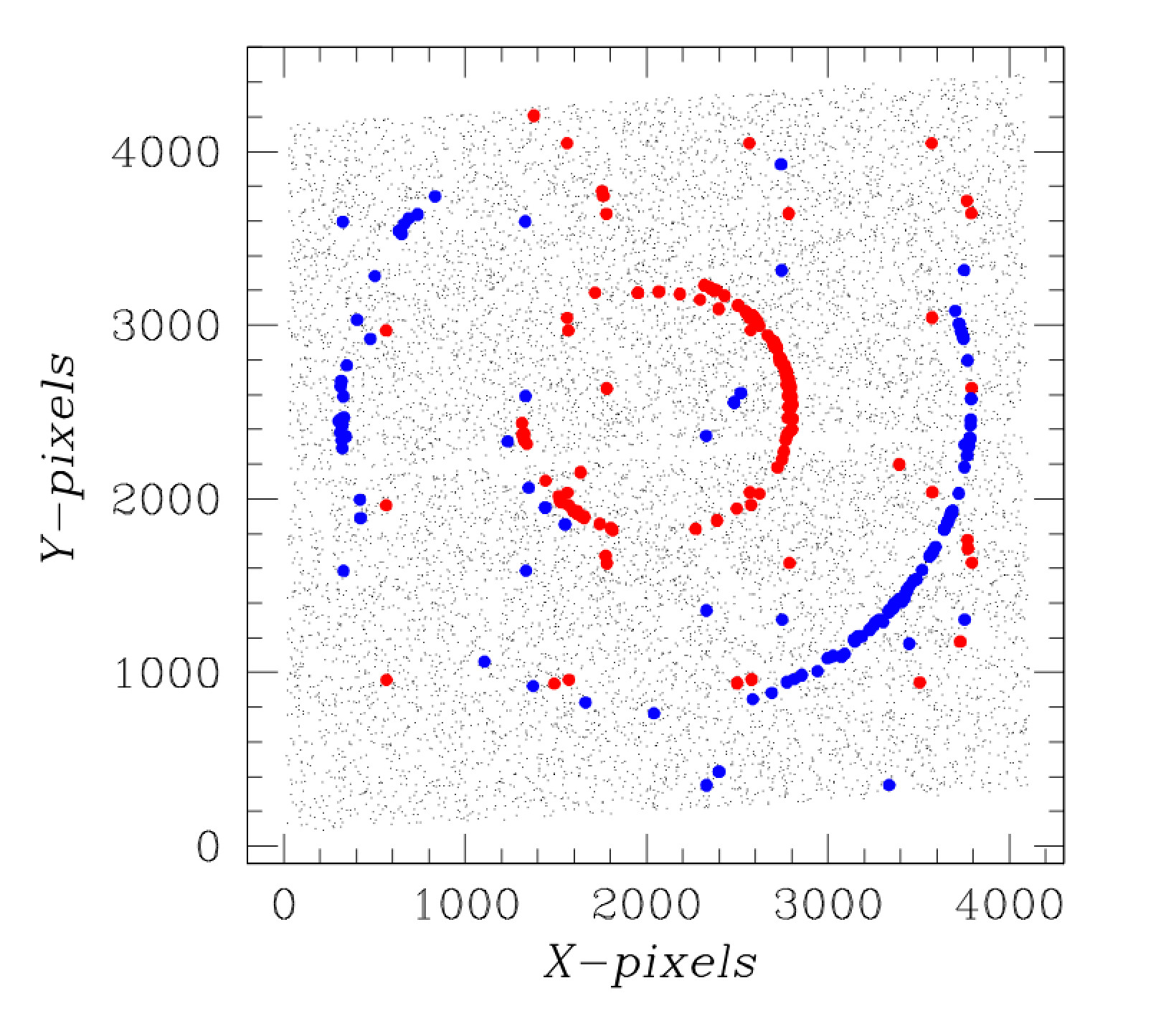}
\caption{Distribution of the locations of two stars on the WFC3/UVIS CCD detector over 175
epochs. Red points correspond to star 290133 with 174 detections, blue points correspond to star 212028 with 149 detections. Red points are more concentrated and that resulted in higher overall accuracy. The background are stars with $m_{F606W}\leq 18.5$.}
\label{fig:distr}
\end{figure}

The observations of $\omega$~Cen were taken with the 
fourth-generation imaging camera WFC3/UVIS 
on-board \hst installed during Servicing Mission 4 in May 2009, with the angular field of view of 2\farcm73 $\times$ 2\farcm73. {The first observation in calibration program
CAL-12094 (PI J.~Kim Quijano)  was taken on 2009-07-15 to  monitor the flat-field
uniformity through the filter F606W at the adopted center} of
$\omega$~Cen (RA=201$\fdg$69283 Dec=$-$47$\fdg$47905).
The following calibration programs CAL-11911 (PI E.~Sabbi), CAL-12094 (PI L.~Petro),
CAL-12339 (PI E.~Sabbi) and the {successive} calibration programs 12353, 12714,
13100, 13570, 14031, 14393, 15000, 15593, 15733, 16413, 16588 (all by PI V.~Kozhurina-Platais) are addressing a variety of specific aspects of the WFC3 instrumental calibration, such as
photometric stability, geometric distortions, and their stability \citep[for the latter aspect, see also][]{Sahlmann:2019ab}. All of these programs used F606W besides some other filters and relatively short exposure times of 35-40 or 60 sec and
a variety of roll angles from $-170\degr$ to 180$\degr$. During the first five years
nearly 50\% of frames had a large 40$\arcsec$ dither. From the view-point
of precision astrometry such dithers could be detrimental due to the partial
overlap. The large variety of roll-angles produced a lower number
of total detections outside the radius of r=80$\arcsec$ from the adopted center. At larger radius and certain roll angles some targets cannot be observed. This is visible in Figure \ref{fig:distr}. Altogether, for this study we selected a total of 175 images spanning 13 years {with the last observation on 2022-01-11.}

The distribution of epochs over each calendar year is not even.
There are two intervals without observations: April~29 to June~3 (35 days) and September~6 to December~12 (97 days). These `holes` are due to the un-schedulability of $\omega$~Cen at those times. Some of the other missing observations are due to failure of guide-star acquisition.  
In some cases, such periodic pattern may spawn spurious one-year signals in the periodogram analysis (Section \ref{sec:periodogram}).

\section{Data reduction}\label{datred}
The chosen \texttt{\_flc-} type images from the \hst Mikulski Archive for Space Telescopes (MAST) are a standard output of the \hst calibration pipeline CALWFC3 with bias and dark subtraction, and flat-fielded on original pixels. The \texttt{\_flc.fits} images are  corrected for the charge-transfer efficiency (CTE) in the \hst pipeline \citep{and21}. The effect of CTE tend to increase with time and can affect the point-spread function (PSF). This can have an impact on the  measurements of stellar centroids and flux. Therefore, the pixel-based corrections is a crucial step towards deriving high-precision measurements of basic stellar parameters, it is implemented in the universal and versatile software routine \texttt{hst1pass} \citep{2022wfc..rept....5A}. It also contains 2-D libraries of empirical PSFs based on specific variations of the PSF due to the \hst thermal breathing and small changes in the \hst focus adjustment to the particular focus level of each exposure \citep{2017MNRAS.470..948A,2018wfc..rept...14A}. 

\texttt{hst1pass} produces the catalog of $X$\&$Y$ positions, 
 the measured flux provided in the instrumental magnitude and \texttt{qfit}, the quality parameter of the PSF fit. We limited our sample to measurements with  0.0$<$\texttt{qfit}$\leq$0.1. The typical distribution of $\texttt{qfit}$ as a function of instrumental magnitude can be found in \citep[Figure~4;][]{pla15}, {which shows two easily-separable populations of artifacts (with larger \texttt{qfit}) and genuine measurements}. Our goal is quality, not completeness, {and the applied constraint only selects the very-best measurements}.  The derived high-precision positions are corrected for geometric distortions \citep{bel11} and \texttt{hst1pass} also simultaneously provided  celestial coordinate RA \& Dec. The WFC3/UVIS geometric distortions are known to be stable over time \citep{2015wfc..rept....2K}.  Nevertheless, the still-ongoing distortion monitoring  is one of the goals in the multi-cycle WFC3 calibration program. 

\subsection{Astrometric registration}\label{reduct}
The output of \texttt{hst1pass} provides a geometrically-correct gnomonic projection for a tiny part of the celestial sphere. We used \gaia Early Data Release~3 (EDR3) \citep{bro21} to provide the reference frame. Preliminary tests showed that {EDR3 source quality is atypical} in the center part of $\omega$~Cen. If we limit the EDR3 sources to the ones with positional error less than 1~mas, then nearly 50\% of them are missing proper
motion and there is a cut-off at $G\sim$18 mag. Clearly, this is a consequence of the crowding and limitations of \emph{Gaia} onboard resources \citep{GaiaCollaboration:2016aa, 2023A&A...669A..55C}. 

Similar to \citet{bel17} we selected the seed-frame \texttt{ibc301qrq\_flc} {{taken on 2010-01-14}}.
If we translate the \gaia celestial coordinates to the epoch 2010.12 and then convert
into gnomonic coordinates and align them to the measured pixel coordinates of the seed frame, the rms of the resulting residuals is $\sim$4.4 mas. There are 440 common stars for this transformation. In contrast, any pair of WFC3/UVIS frames with close epochs
generates an rms of $\sim$0.7 mas. Such disparity may imply that the EDR3 positions near the center of $\omega$~Cen have substantially lower accuracy then expected. For this reason we decided to use the EDR3 positions only for aligning the seed-frame to the RA and Dec axes and to derive the WFC3/UVIS pixel scale in the F606W filter. 

To align the seed frame it had to be rotated by 149$\fdg$751 around the location
of star 227418 \citep[Table~7][]{bel17}. At this stage we decided to
use the \citet{bel17} numbering system to designate our sources since it covers
the same area. Also, the availability of celestial coordinates in both data-sets make it easy to cross-identify the sources.
Our estimate of the WFC3/UVIS pixel scale is 39.770 mas and is not corrected
for the velocity aberration. As noted by \citet{bel11, and07} the effect of velocity aberration changes only the pixel scale. In this sense, our pixel scale is not absolute. 

Next, all frames were converted into pixel-based coordinates of the seed-frame using a least-squares minimization. The number of common stars in the magnitude range $15.4<{\rm F606W}<17.5$ span from 1093 to 2434 depending on the dither pattern and the \hst roll angle. If the pixel coordinates are free of geometric distortions, then a linear 3-term polynomial should be sufficient to align any pair of frames. Our analysis of the residuals yielded the following results: 
\begin{itemize}
    \item The rms of the residuals is correlated with the epoch of observations. Early epochs produce low rms at $\sim$0.015~pix whereas the last epochs have values as high as $\sim$0.22~pix. The significant internal velocity dispersion near the center
of $\omega$~Cen at 0.75~mas~yr$^{-1}$ \citep{and10} is the main reason this increase. Such patterns are not only detrimental to the astrometric
accuracy but may also impact the prospects of finding accelerations.

\item Initial solutions using only three linear terms revealed a small (up to 0.05 pix) 
but persistent quadratic pattern in the residuals, mainly along the $Y$-axis. Unfortunately, we do not know whether the seed-frame itself is free from quadratic markings. There are a handful of cases with nearly perfect distribution of residuals but not necessarily at the same \hst roll angle. 

\item About a dozen frames show atypical box-shaped distributions of residuals. Since
such frames appear close in time to typical frames, we suspect that sometime
the guiding of \hst was not perfect, especially for some short exposures
used in this calibration program.
\end{itemize}

The solution to these issues is a three-pronged approach: 
\begin{itemize}

\item Subtract from all $X$\&$Y$ pixel-positions the effect of proper motion relative to the epoch of the seed-frame. {We
subtracted the effect of the known relative proper motion from each $X$\&$Y$ position at every epoch relative to the epoch of the seed-frame. In practice, the known proper motion is multiplied by the time difference in years and the result is subtracted from $X$\&$Y$. As the result, all positions have zero proper motion up to its formal precision. Then, these positions are averaged and the mean is subtracted. In essence, our approach levels the positional accuracy over all epochs. This can be seen in Figure \ref{fig:4_accl} in the even scatter of residuals in time.}  At each epoch we calculate the amount of this effect and convert it into pixels using exact scale at 40 mas per pixel and then rotate around the adopted zeropoint at $X_c$=2073.19 and $Y_c$=2292.38 pix using the \hst position angle $PA\_V3$.  Since the proper motion is strictly vectorial, it cannot affect the inferred acceleration. The opposite, however, is true: acceleration can bias the inferred proper motion.

\item The modified pixel coordinates $X_m$, $Y_m$ are then translated into the system of the seed-frame positions by a least-squares regression using a polynomial model with 6 terms (linear and quadratic) for both axes \citep[e.g.,][]{bel14}. The output are epoch positions $X_f$, $Y_f$  free of proper-motion effects to the level of their accuracy. For sources of particular interest, e.g.\ with likely acceleration signals, we also subtracted a small local offset in both coordinates by using all sources within $\sim$~50 pixels around the target and at all epochs. Then the averages of these offsets are applied to all epochs. 

\item A total of 17 epochs are excluded from the final analysis, the recalculation of proper motion, and the search for acceleration. It is most likely that at these epochs \hst experienced jitter in its pointing stability due to the degrading gyro \citep{2018wfc..rept....7A}. There is no sharp divide between the normal jitter ($\sim$ 11~mas) and the increased ones. We note that such \hst calibration frames may not be science grade but still can be used for engineering purpose. Hence they are in the MAST.
\end{itemize}

\section{LONG-PERIOD ACCELERATIONS}\label{long_accel}
\begin{figure}[h!]
\includegraphics[trim={2cm 4.0cm 2cm 0.5cm}, clip, width=\columnwidth]{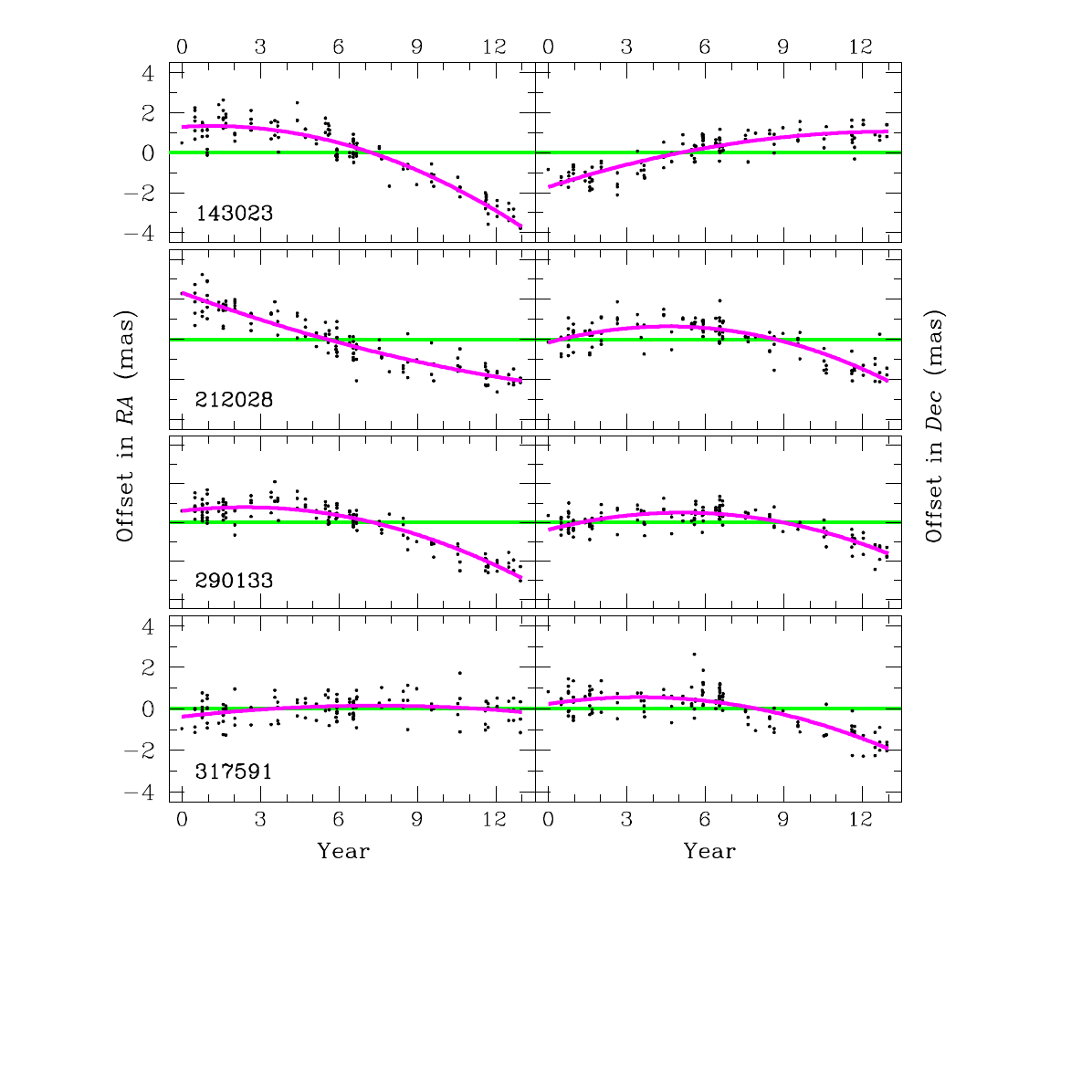}
\caption{Four cases of long-period acceleration. Magenta lines show quadratic
least-square fits. Green lines indicate zero offset. If a star lacks acceleration, all measured $XY$ (black points) would cluster around a straight line in both axes. As an example of borderline acceleration is star 317591 in $X$-axis. This star would be rejected if not the prominent acceleration in the $Y$-axis.
\label{fig:4_accl}}
\end{figure}

The new astrometric catalog includes 162,604 sources down to $m_{F606W}\simeq 21.5$. We examined the parameter space so that it is optimal for the detection of accelerations. {The suitable  limits appear to be at $m_{F606W}$ $16.1-20.0$ mag and no less then 120 epochs evenly spread over 13 years.} There are 22,466 such sources. Each set of $X_f$,$Y_f$ then is used to calculate small  additional proper motion and acceleration which can be approximated by the quadratic term in each axis, $X$$^{2}$ and $Y$$^{2}$. Thus, the output of the least-squares minimization yields two formal accelerations along each axis, named $\xi^{-2}_{X}$ and $\xi^{-2}_{Y}$ (see Table~\ref{tab:info}). The crucial parameter is the ratio the acceleration term and its formal error, designated as $Q_X$ and $Q_Y$. This ratio (ignoring the sign) should be at least $\sim$5 in one or both axis (see $Q_X$ and $Q_Y$ in Table~\ref{tab:info} and Figure~\ref{fig:4_accl}). With this criterion we find four clear cases of astrometric acceleration discovered in a globular cluster. 

Using a smaller threshold of 3-5 on the $Q$-parameter generates $\sim$400 candidates, but these are heavily polluted by impostors. A typical example is an offset or unusual spread at some narrow bunch of epochs in the same \hst program. Generally, our search is limited by the known noise-floor of geometric-distortion corrections at $\sim$0.3~mas \citep{bel11}. 

In addition, the long-period acceleration of source 233697 was detected in the independent periodogram analysis of the astrometric time-series, see Section \ref{sec:periodogram}. Among the astrometric accelerations this source has the shortest period at 8.8 yrs. Due to the complete orbit and more, the quadratic term will diminish or may even flip its sign.   

{We inspected the F606W photometric timeseries and found them to be stable with robust scatter estimates \citep{Lindegren:2012aa} of between 14 and 42 mmag, without noticeable correlations with the acceleration signals.}

\begin{figure}[ht!]
\includegraphics[trim={0.5cm 1cm 0.5cm 0.5cm}, clip, width=\columnwidth]{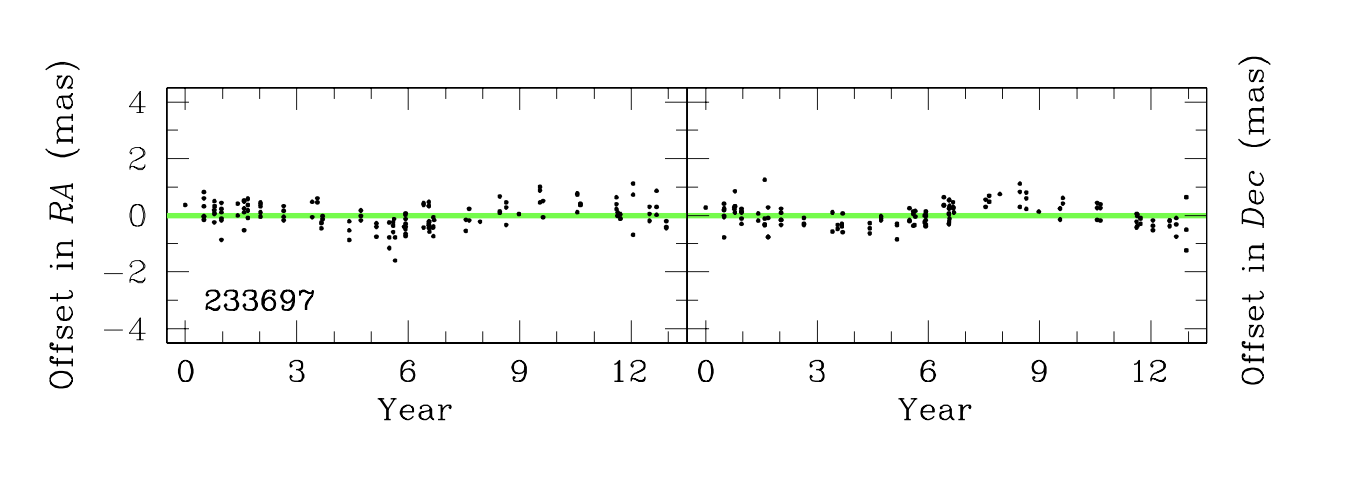}
\caption{Star 233697 has a low acceleration signal, indistinguishable from normal stars. Its Keplerian motion was detected in the periodogram (Figure \ref{fig:member_periodograms}). Green line indicates zero offset, although a careful inspection shows small outbalances within 1~mas.
\label{fig:1_accl}}
\end{figure}

 \begin{table}
  \centering
  \caption{Astrometric binaries {discovered in this work}: astrometry {at the Gaia DR1 reference epoch J2015.0 (from \citealt{bel17})} and photometry of the visible components and acceleration parameters. $m_V$ and $m_I$  are HST WFC3 Vega magnitudes in F606W and F814W filters; proper motions $\mu_X$, $\mu_Y$, and their errors $\sigma_{\mu_X}$, $\sigma_{\mu_Y}$ in mas~yr$^{-1}$; $N_{\rm frame}$ number of frames (epochs)  followed by accelerations and their significance. {The last column indicates our designation of these newly-discovered binaries.}}
  \label{tab:info}
\begin{tabular}{lcccc}
\hline
\hline

 ID  &  $m_{V}$ & $m_{V}-m_{I}$ & RA (deg) & Dec (deg) \\
\hline
143023 & 17.056 &0.855 & 201.6675417 & -47.4739278 \\
212028 & 18.073 &0.610 & 201.7154333 & -47.4673472 \\ 
233697 & 17.398 &0.650 & 201.6714500 & -47.4852972\\
290133 & 18.148 &0.591 & 201.7047500 & -47.4809222 \\
317591 & 18.811 &0.633 & 201.7239958 & -47.4788750\\

\noalign{\smallskip}
\noalign{\smallskip}
\hline
\hline
$\mu_X$  &  $\mu_Y$ &   $\sigma_{\mu_X}$ &  $\sigma_{\mu_Y}$ & $N_{\rm frame}$\\
\hline
    -2.350 &   5.653   &  0.017 & 0.014 &  135\\ 
    \phn0.434 &  -0.171   &  0.013 & 0.018 & 134\\  
    \phn0.951 &   -0.023  &  0.011 & 0.011 & 139\\ 
    -0.407 &   0.233   &  0.013 & 0.013 & 157\\ 
    \phn0.882 &   0.270   &  0.013 & 0.016 & 132\\  
\noalign{\smallskip}
\noalign{\smallskip}

\hline
\hline
$\xi^{-2}_{X}$ & $\xi^{-2}_{Y}$ & $Q_X$ & $Q_Y$ & Designation\\
\hline
-0.0359 & -0.0159 & 10.5 &  5.4 & HSToC-1 AB\\
 0.0109 & -0.0387 &  3.0 & 11.4& HSToC-2 AB\\
 \nodata & \nodata & \nodata & \nodata & HSToC-3 AB\\
-0.0329 & -0.0343 & 12.9 & 14.7& HSToC-4 AB\\
-0.0098 & -0.0274 & 2.8 &  7.7& HSToC-5 AB\\
\hline
\end{tabular}
\end{table}

\section{Masses of visible components from stellar evolutionary models}\label{sec:isoc}
\begin{figure}[h!]
\includegraphics[trim={0.5cm 0.5cm 0.5cm 0.5cm}, clip, width=\columnwidth]{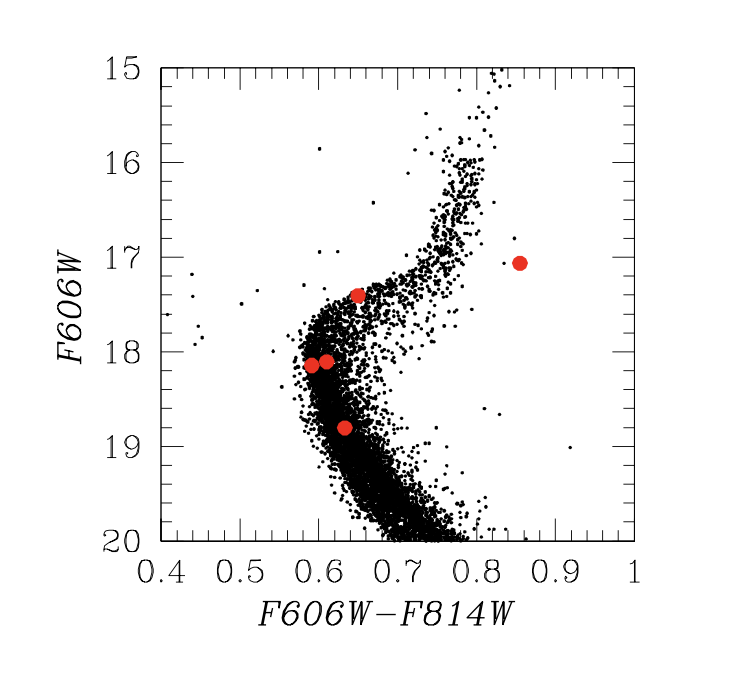}
\caption{Observational CMD of $\omega$~Cen  in the \hst photometric Vega system through WFC3/UVIS F606W and F814W filters \citep{bel17}. {A randomly-chosen subset of $\sim$10\%  of all stars is shown. Red circles show the locations of the astrometric binaries. The reddest of these is the field binary \onefour}.
\label{fig:obs_cmd}}
\end{figure}

Three of the cluster members with high acceleration are clearly main sequence stars located just below  the turn-off of $\omega$ Cen, see Fig.~\ref{fig:obs_cmd}, and the fourth one (ID=233697) is on the sub-giant branch.

In the absence of any information other than the photometry, their mass can only be estimated via theoretical isochrones \citep{massreview}. For instance, one can simply select the isochrone with the most appropriate value of age and metallicity, then shift it to the appropriate distance modulus and foreground reddening and then read the stellar mass from the isochrone that is closest to the observed star in the color-magnitude diagram (CMD). 

In the case of $\omega$~Cen this procedure is complicated due to the wide range of metallicity  \citep[see e.g.][]{frinchaboy02}. Indeed, if we take a 12-Gyr isochrone with the mean metallicity of $\omega$ Cen, namely [Fe/H]$=-1.53$, and relocate it to the distance of 5.24 kpc \citep{sol21} and redden it by $E(B-V)=0.132$~mag \citep[cf.][]{bono19}, then the high-acceleration stars clearly fall to the red, meaning that they are probably more metal rich.
 
\begin{figure}[h]
\includegraphics[trim={2cm 0cm 3cm 0cm}, clip, width=\columnwidth]{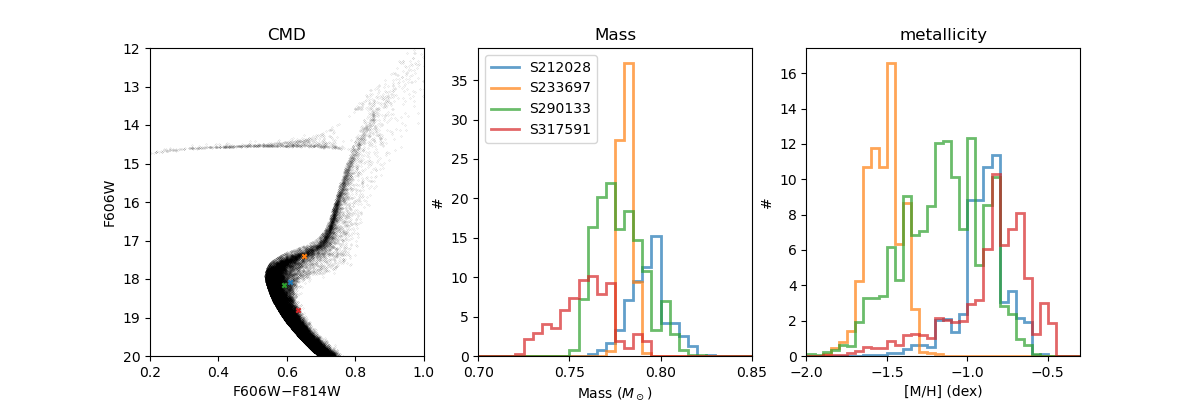}
\caption{
Left panel: a simulated CMD of $\omega$~Cen (black tiny dots), built to estimate the masses and metallicities of the four visible components with high acceleration (coloured  dots). 
{The central and right panels show the probability density functions for both mass and metallicity for these stars, as derived from the stars with similar color--magnitude position in the simulation (see text for details).}
} 
\label{fig:stars_cmd}
\end{figure}

In order to create a more realistic model for these stars, we simulated a cluster composed of single stars only, with an age of 12 Gyr and {using the entire distribution of [Fe/H] derived by \citet{frinchaboy02}. We note that their metallicity distribution function is based on photometry with Washington $M$, $T_2$ and DDO51 filters, but that it was confirmed by the high-resolution spectroscopy from APOGEE \citep{meszaros21}}. This simulation contains a tail of metal-rich stars extending far enough to the red  and cover our four accelerating stars (see left panel in Fig.~\ref{fig:stars_cmd}). 

\begin{table}[h!]
  \centering 
  \caption{Astrophysical parameters of the visible {components}. Masses were estimated on the basis of the sources' positions in the CMD.}
  \label{tab:astro_parameters} 
\begin{tabular}{cccc}
\hline
\hline
 ID & $M_\odot$ & [Fe/H] \\
\hline
143023 & \multicolumn{2}{c}{See Appendix \ref{sec:foreground}}\\
\multicolumn{3}{c}{If $d=5.24\pm0.11$ kpc:}\\
212028  & $0.794_{-0.012}^{+0.013}$ &  $-0.82_{-0.18}^{+0.12}$ \\
233697  & $0.784_{-0.004}^{+0.005}$ &  $-1.38_{-0.10}^{+0.12}$ \\
290133  & $0.773_{-0.013}^{+0.014}$ &  $-1.14_{-0.32}^{+0.26}$ \\
317591  & $0.750_{-0.017}^{+0.014}$ &  $-0.75_{-0.38}^{+0.16}$ \\
\multicolumn{3}{c}{If $d=5.426\pm0.047$ kpc:}\\
212028  & $0.793_{-0.010}^{+0.010}$ &  $-0.78_{-0.14}^{+0.12}$ \\
233697  & $0.780_{-0.002}^{+0.004}$ &  $-1.48_{-0.10}^{+0.10}$ \\
290133  & $0.776_{-0.012}^{+0.014}$ &  $-0.90_{-0.12}^{+0.14}$ \\
317591  & $0.760_{-0.017}^{+0.013}$ &  $-0.66_{-0.31}^{+0.15}$ \\
\hline
 \end{tabular}
\end{table}

{To estimate the masses of the four visible components with high acceleration, we proceed as follows: For every star in the simulation with color $c$ and magnitude $m$, a weight $w_i$ is assigned as 
\begin{equation}
w_i = \exp[-(c_i-c_0)^2/\sigma^2-(m_i-m_0)^2/\sigma^2]    
\end{equation}
where $(c_0,m_0)$ is the color-magnitude position of the visible components, and $\sigma=0.02$ mag is the typical photometric error for both color and magnitude. These weights reflect the likelihood that the color-magnitude location of the observed stars derive from the theoretically-predicted location plus a Gaussian distribution of photometric errors. These weights are used together with the masses and metallicities of the simulated stars, to build the probability density functions (PDF) for both mass and metallicity. These functions are depicted in the central and right panels of Fig.~\ref{fig:stars_cmd}. From these PDFs, we then derive the median and 68\% confidence ($1\sigma$-equivalent) intervals for both mass and metallicity. Their values are listed in Table~\ref{tab:astro_parameters}. } 

{Table~\ref{tab:astro_parameters} initially presents the values obtained by assuming the standard reference to the distance of $\omega$~Cen \citep[5.24$\pm0.11$ kpc;][]{sol21} based on \gaia EDR3 parallaxes. The error in distance modulus in this case is of just 0.004~mag, which is one fifth of the error we already assume in our derivation of the masses and metallicities, and hence not able to significantly affect our estimations. However, there are alternative values to $\omega$~Cen distance that are worth exploring.
The review of GC distances by \citet{baumgardt21}, for instance, provides a $5.426 \pm 0.047$ kpc distance; these authors argue that the \gaia parallaxes have small-scale correlated  errors and systematic biases but this is not well yet understood in EDR3, particularly for $\omega$~Cen. The bottom part of Table~\ref{tab:astro_parameters} presents the mass and metallicity estimates that are obtained with this distance value. As it can be appreciated, changes in the mass values are smaller than $0.01 M_\odot$.}

{En passant,} we note that the location of our high-acceleration stars on the simulated CMD indicate that the metallicities of these stars are higher than the mean value of $\omega$~Cen [Fe/H]$=-1.53$ (see Table~\ref{tab:astro_parameters}). 

\section{Periodograms of astrometric time-series}\label{sec:periodogram}
To search for deviations from the linear astrometric motion expected for a single source, we processed the two-dimensional astrometric timeseries as follows. For every source we subtracted the average position from the timeseries and fitted a four-parameter linear model corresponding to two positional offsets and two proper motions. {This analysis is therefore independent of the amount of proper-motion displacement subtracted in the pre-processing and analysis discussed in Sect.~\ref{reduct}.} We omitted the parallax term because we do not expect to measure the parallax signal for cluster members, due to the inherently relative nature of our astrometric measurements.

\begin{figure}[ht!]
\includegraphics[width=\columnwidth]{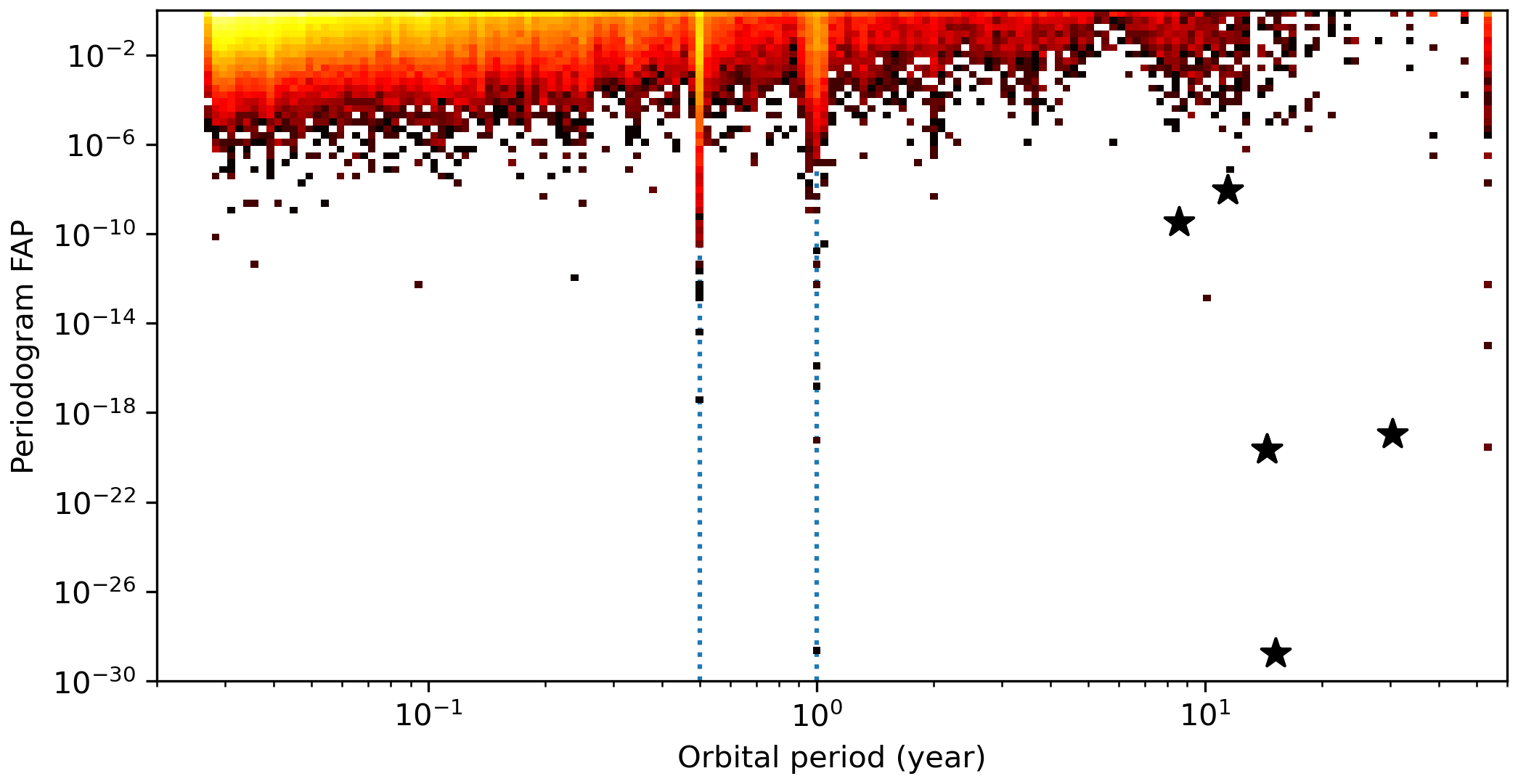}
\caption{Density histogram of FAP as a function of peak period for the $\sim$160\,000 sources for which the periodogram could be computed. The two vertical lines are at periods of 0.5 and 1 year. The black stars indicate the location of the five sources in Table \ref{tab:info}.} 
\label{fig:period_fap}
\end{figure}

We then computed the periodogram of the fit residuals using the \texttt{kepmodel} package \citep{2022A&A...667A.172D}. The period and false-alarm probability (FAP) of the highest periodogram peak were retained and are shown in Figure \ref{fig:period_fap}. The FAP indicates the probability that the peak is caused by random noise, i.e.\ a small FAP indicates the presence of periodicity in the time-series. The lower and upper boundaries of the periodicity search were 10 days and 20\,000 days, respectively.

Most sources exhibit insignificant periodogram peaks with large FAP, and we performed detailed analysis for all sources with small FAP. A few sources show significant signals with periods of one year or half a year. These are likely foreground sources for which measurable parallax offsets are present.  The list of sources with FAP $<10^{-10}$ and periods around half a year is 420482, 414263, 396157, 421049, 421381, 427775, 428247, 428253, 428571, 434388, 412665, 421379, 421520, 427880, 396159, 421074. The list of sources with FAP $<10^{-10}$ and periods around one year is 247164, 276969, 109223, 87685, 387712, 116672, 107882, 312186. {As a by-product of our analysis, we have therefore identified 24 sources that probably are foreground sources and not members of the Omega~Cen cluster.}

The pile-up at long periods is caused by the upper boundary of the period search. Some of these peaks have small FAP but visual inspection of the time-series revealed systematic offsets and/or few measurements in the astrometric timeseries. The same applies to the few short-period ($P<100$ day) sources and to the small-FAP source at $P\sim10$\,yr visible in Figure \ref{fig:period_fap}.

Finally, we identified several sources with long-period periodogram peaks and small FAP whose origin is genuine Keplerian motion. 

\subsection{Constraints on the orbital parameters and masses of the companions}
\begin{figure}[ht!]
\plotone{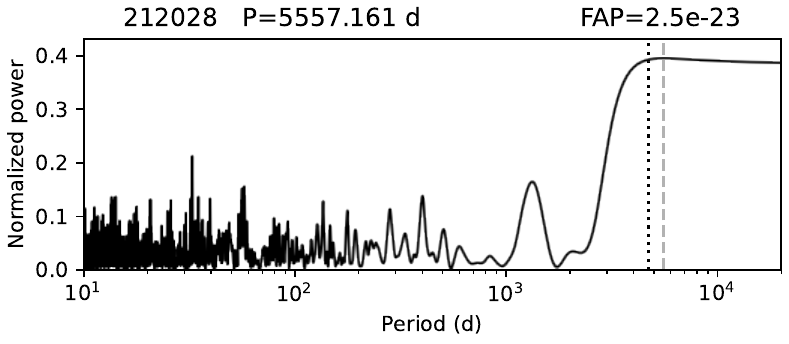}
\plotone{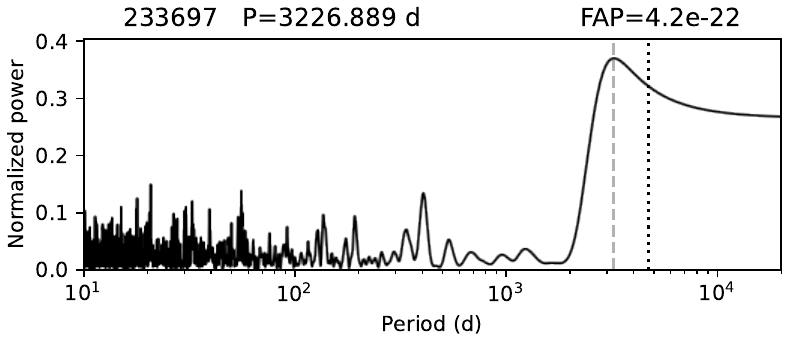}
\plotone{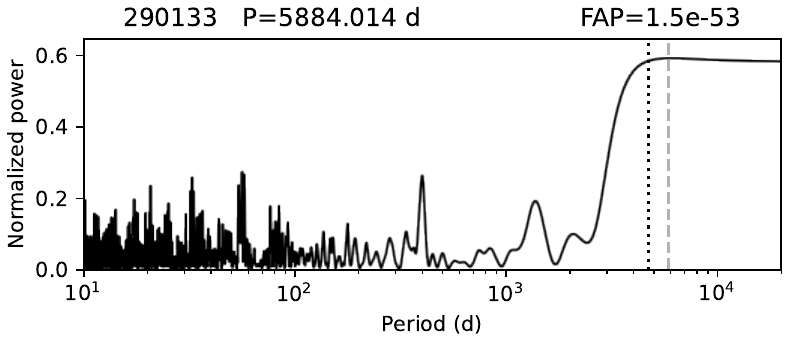}
\plotone{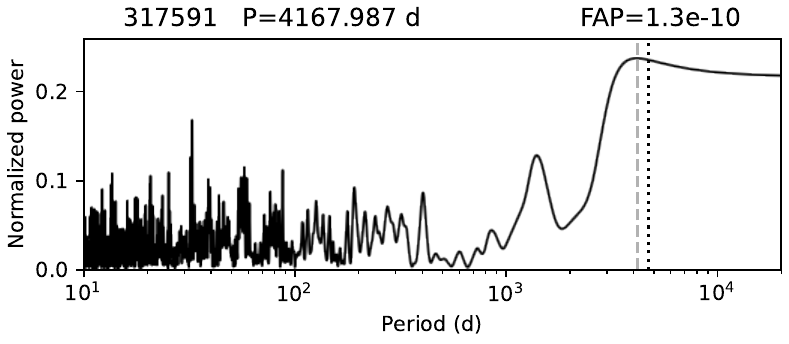}
\caption{Residual periodogram of the four-parameter model for 212028 (top), 233697 (second), 290133 (third), and 317591 (bottom). The vertical grey dashed line shows the location of the highest peak, whose period and FAP are indicated in the panel title. The black dotted line corresponds to the observation timespan for this source.} 
\label{fig:member_periodograms}
\end{figure}

For the sources identified in the previous section on the basis of their small periodogram FAP, we attempted to derive constraints on the possible orbital parameters. One source (143023) turned out to be a foreground binary and is discussed in Appendix \ref{sec:foreground}. The periodograms of the astrometric timeseries residuals for the remaining four sources are shown in Fig.\ \ref{fig:member_periodograms}. All of them except 233697 show a dominant peak at roughly the observation timespan with a large tail towards longer periods. This is typical for astrometric data that cover only a small or moderate portion of the astrometric orbit. This is also evident when inspecting the timeseries in Fig.\ \ref{fig:4_accl}. 

For source 233697{, whose timeseries is shown in Fig.\ \ref{fig:1_accl}}, the periodogram peak is at $\sim$ 3230 days (8.85 yrs) with an observation timespan of $\sim$4730 days (12.96 yrs), indicating that more than one orbital revolution has been observed.

We analyse the data of these sources under the assumption that the non-standard motion is caused by a single dark companion in a Keplerian orbit about the visible source. 
In three out of four cases, the observation time-span is too short to derive {tight} constraints on the orbital parameters and we therefore attempt to determine the range of possible configurations and their astrophysical interpretations. 

For every source we use \texttt{kepmodel} to fit the data with the initial four-parameter model plus a 7-parameter Keplerian model with initial parameters estimated from the periodogram. {The orbit-fitting results are therefore independent of the amount of proper-motion displacement subtracted in the pre-processing (Sect.\ \ref{reduct}).} We do not impose priors on any of the parameters but we set an upper limit of 0.3 to the eccentricity {for 143023 and 317591}. {For the remaining sources} that limit is 0.95. This is to avoid high-eccentricity solutions which often are preferred in orbit-fitting results using incomplete orbital coverage or otherwise sub-optimal sampling. We assigned a uniform uncertainty of 0.3 mas to every individual astrometric measurement, which is motivated by the distortion-correction noise floor. This choice is validated by the typical value of an additional jitter term, which we included in the model.  The model fitting is always performed on the individual-frame data, hence no binning is applied. For visual display, however, we compute normal points as averages of individual frames grouped in time.

We fit the combined model with a standard {maximum-likelihood} algorithm and report a few of the best-fit parameters in Table \ref{tab:orbit_parameters}, which has some caveats:
\begin{itemize}
    \item The algorithm tends to converge towards the smallest period compatible with the data set. However, the true period of the system may be much longer which influences the estimated parameters of companion. 
    \item For {two} sources we applied the $e<0.3$ constraint. The true eccentricities may be larger than 0.3, which {can} imply larger semi-major axe and, therefore, companion mass estimates. {Table \ref{tab:orbit_parameters} also shows the solution parameters for larger eccentricities. For 143023 a larger eccentricity results in a larger semimajor axis and longer period, but the companion mass is nearly unchanged. For 317591 a larger eccentricity results in a longer period and larger companion mass, but the semimajor axis is nearly unchanged.}
    \item When inferring the companion mass, we assumed one single dark companion. This assumption may not be justified, which we discuss {below}. 
    {The alternative scenarios include a pair of quasi-dark companions or a companion object that is not dark but contributes significant light. In the latter case, the mass of that companion has to be even larger than what we inferred.}
\end{itemize}
The uncertainties of orbital period $P$, eccentricity $e$, and semi-major axis of the primary $a_1$ (Table \ref{tab:orbit_parameters}) correspond to the formal uncertainties returned by the minimisation routine and are indicative only. The companion mass $M_2$ was estimated from the orbital parameters and an average of the primary-mass estimates $M_1$ in Table~\ref{tab:astro_parameters}\footnote{The systematic primary-mass uncertainty has insignificant influence on the results.}, with uncertainties propagated using Monte-Carlo resampling.

{Since several of these orbits are not fully covered by our data and both proper motions and orbital parameters are free parameters in the fitted model, it can happen that proper motion is mistaken for orbital motion and vice versa. This is however not the case for our solutions.}
We verified that the residual proper motion determined as part of the orbital fit is {comparable to or smaller than} its uncertainties. 
This is important to safeguard against potential instabilities in the fitting routine where a spurious proper motion and long-period Keplerian offsets compensate each other.

\begin{table*}[h!]
  \centering 
  \caption{Preliminary estimates of some Keplerian parameters and the masses of the companions. {These were obtained under the assumption of a single and dark companion and with a prior on eccentricity for sources \onefour/143023 and \threeone/317591.} Formal uncertainties from the minimisation routine are given only for orientation. {$M_1$ is derived from Table~\ref{tab:astro_parameters}. The foreground source \onefour/143023 is discussed in Appendix \ref{sec:foreground}}. {The second part of the table shows indicative results for \onefour/143023 and \threeone/317591 when relaxing the eccentricity prior and allowing for values of $\sim$0.5 and $\sim$0.8. The $K_\mathrm{RV}$ is the estimated radial-velocity amplitude.}}
  \label{tab:orbit_parameters}
\begin{tabular}{lrrrrrr}
\hline
\hline
Designation / ID & Period & $e$ & $a_\mathrm{photo}$&$M_1$ &$M_2$ & $K_\mathrm{RV}$\\
 & (yrs) &  & (mas) & ($M_\odot$) & ($M_\odot$) & km/s \\
\hline
\onefour\ / {143023} & $19\pm8$  & $<0.3$ & $0.9 \pm 0.3$ &  $0.68\pm0.03$ & $0.15^{+0.06}_{-0.05}$& $\sim0.9$\\
\twoone\ / 212028 & $15\pm3$  & 0 & $0.91\pm0.27$ &  $0.80\pm0.01$ & $1.25^{+0.87}_{-0.70}$& $\sim9$\\
\twothree\ / 233697 & $8.8 \pm 0.5$  & $0.2 \pm 0.2$ & $0.47 \pm 0.05$ &  $0.784\pm0.005$ & $0.77\pm0.14$& $\sim6$\\ 
\twonine\ / 290133 & $13 \pm 2$  & $0.3\pm0.2$ & $0.86\pm0.08$ &  $0.78\pm0.01$ & $1.36^{+0.36}_{-0.26}$& $\sim10$\\
\threeone\ / 317591 & $11.5\pm2.0$  & $<0.3$ & $0.53\pm0.07$ &  $0.75\pm0.02$ & $0.71^{+0.23}_{-0.16}$& $\sim7$\\
\hline
\hline
\onefour\ / {143023$^{*}$} & $\sim22$ & $\sim0.50$ & $0.9 \pm 0.3$ &  $0.68\pm0.03$ & $0.19^{+0.04}_{-0.04}$& $\sim1$\\ 
\onefour\ / {143023$^{*}$} & $\sim44$  & $\sim0.80$ & $1.4 \pm 0.9$ &  $0.68\pm0.03$ & $0.19^{+0.23}_{-0.09}$& $\sim1$\\ 
\threeone\ / {317591$^{*}$} & $\sim12$  & $\sim0.50$ & $0.6 \pm 0.1$ &  $0.75\pm0.02$ & $0.7^{+0.4}_{-0.2}$ & $\sim8$\\
\threeone\ / {317591$^{*}$} & $\sim14 $  & $\sim0.80$ & $0.6 \pm 0.3$ &  $0.75\pm0.02$ & $0.8^{+1.2}_{-0.4}$ & $\sim11$\\

\hline
\end{tabular}
\end{table*}

Although the systematic and formal uncertainties are significant, we can categorize the four binaries into two groups: 233697 and 317591 are probably systems with similar component masses, whereas the solutions for 212028 and 290133 indicate companions that appear to be significantly more massive than the primary star, but yet might be much dimmer.

The secondary masses in Table \ref{tab:orbit_parameters} were derived under the assumption that the companion is dark, i.e.\ the positions of the system's photocentre and the primary coincide. The semimajor axis of the photocentre motion $a_\mathrm{photo}$ that we determine directly is then equal to the semimajor axis of the primary’s barycentric motion $a_1=a_\mathrm{photo}$. If the companion contributes light, the photocentre motion gets diluted and $a_1>a_\mathrm{photo}$ which implies an even larger mass of the companion. 
{The scaling between a binary’s photocentre and barycentre orbit size is determined by the fractional mass $f = M_2/(M_1 + M_2)$ of the components and their fractional luminosity $\beta = L_2/(L_1 + L_2)$ \citep{Heintz:1978yq}:
\begin{equation}\label{eq:photocenter_orbit}
    \bar a_\mathrm{photo} = \bar a_\mathrm{rel} (f - \beta),
\end{equation}
where $\bar a_\mathrm{photo}$ is the photocentre orbit size in linear units and $\bar a_\mathrm{rel}$ is the semimajor axis of the relative orbit in linear units. Using Kepler's third law in SI units this can be written as 
\begin{equation}
    \bar a_\mathrm{photo} = \sqrt[3]{G P^2/(4 \pi^2)} \sqrt[3]{M_1+M_2} (f - \beta)
\end{equation}
and therefore
\begin{equation}\label{eq:photo}
    a_\mathrm{photo} \propto \bar a_\mathrm{photo} \propto \sqrt[3]{M_1+M_2} \left( \frac{M_2}{M_1 + M_2} - \frac{L_2}{L_1 + L_2}\right).
\end{equation}
From Eq.\ \ref{eq:photo} it is clear that to maintain the observed $a_\mathrm{photo}$ constant between the dark companion limit (where $L_2=0$ and $a_1=a_\mathrm{photo}$) and a luminous companion with $L_2>0$, the companion mass $M_2$ has to increase. 
}

With the data available to us we cannot exclude that the companions contribute light, but if they do it implies that they are fainter and at the same time more massive than the primary.

{Two configurations that could satisfy these constraints for \twoone and \twonine, which have the largest estimated companion masses, are (a) the companion is a tight pair of white dwarfs (see Section \ref{sec:Discuss}) or (b) the companion is a tight pair of main-sequence stars. We simulated the latter configuration by generating synthetic triple systems that have the same astrometric signature in terms of period but are composed of a primary with properties from Table \ref{tab:astro_parameters} and a twin-binary companion of the same metallicity with individual masses between $\sim$0.4--0.8\,$M_\sun$. Using the isochrones discussed in Section \ref{sec:isoc} and Eq.\ \ref{eq:photocenter_orbit} we found that the corresponding photocenter orbit size is largest when the total companion mass is $\sim$1\,$M_\sun$, but that it remains $0.2-0.25$ mas smaller than the measured orbit size. This scenario therefore seems excluded for \twonine and unlikely for \twoone.}

\subsection{Orbital motion of 290133/\twonine}
\begin{figure}[ht!]
\includegraphics[trim={0 0 0cm 0.5cm},clip, width=\columnwidth]{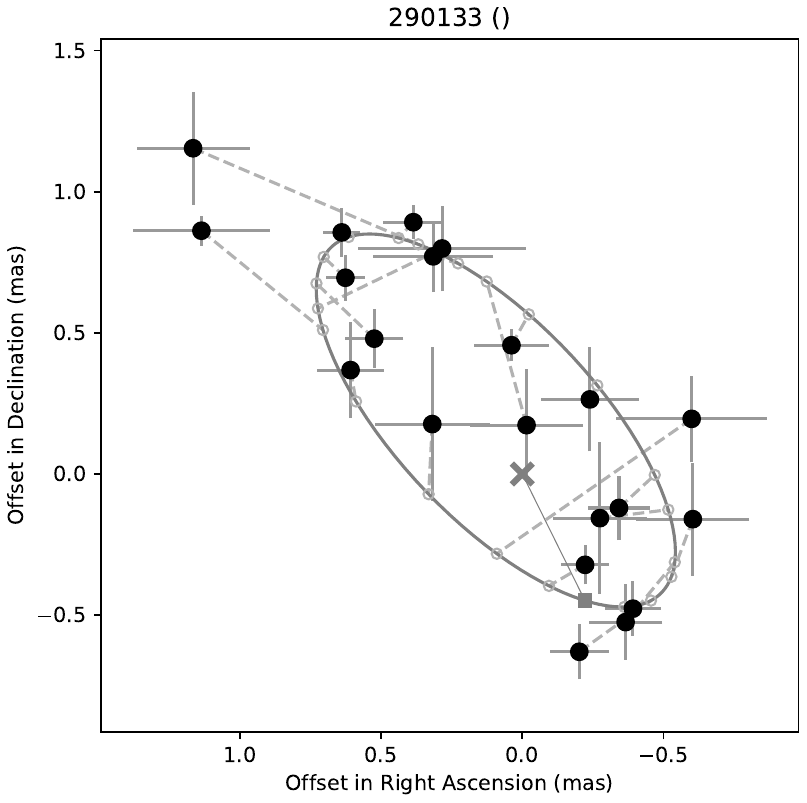}
\caption{Orbital motion of \twonine/290133 in the sky after subtracting the standard 5-parameter model. The grey curve corresponds to the best-fit orbital parameters. Uncertainties represented as error-bars were estimated from the dispersion of the data accumulated in the normal point. Dashed lines connect observed and computed locations.}
\label{fig:290133_orbit_sky}
\end{figure}

\begin{figure}[ht!]
\includegraphics[trim={0 0 0cm 0.5cm},clip, width=\columnwidth]{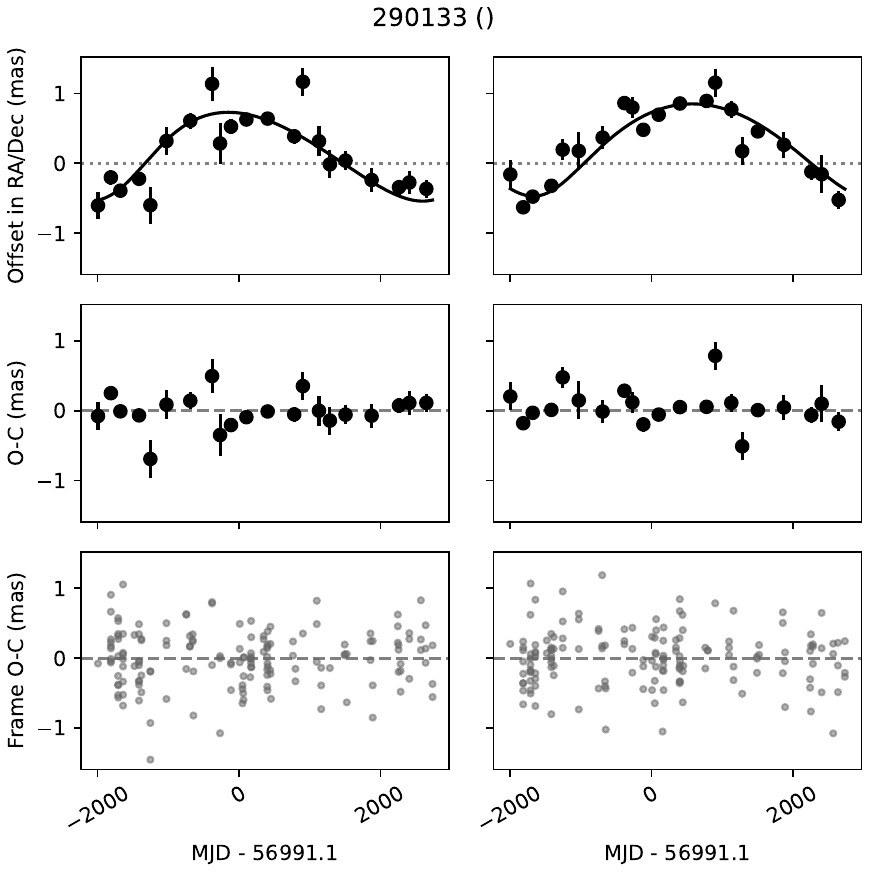}
\caption{Orbital motion of \twonine/290133 after subtracting the standard model (top panels) as a function of time. The final residuals of the full-model fit are shown in the middle and bottom panels. The top and middle panel show normal points computed by combining individual frames, whereas the bottom panels show individual frame data. Left panels are in RA direction and right panels are in Declination.}
\label{fig:290133_orbit_time}
\end{figure}

The source 290133 stands out as having the largest estimated companion mass.
We show the orbital motion of 290133 in the plane of the sky in Figure \ref{fig:290133_orbit_sky} and as a function of time in Figure \ref{fig:290133_orbit_time}. For the sake of better visualisation only, we computed averages over time-bins of $\sim$100 days (normal points). We stress that these figures show only one possible realisation of the orbital parameters, the actual orbit may look different. However, these visualisations are useful to ascertain the  Keplerian orbital motion.

In Figure \ref{fig:290133_orbit_time} it can be appreciated that the excess signal after subtracting parallax and proper motion has a standard deviation of 0.51 mas (both axes, computed on normal points) which is reduced to 0.25 mas when including the Keplerian model. The equivalent figures for the other sources are shown in Figures \ref{fig:212028_orbit_time}, \ref{fig:317591_orbit_time}, and \ref{fig:233697_orbit_time} of the Appendix.

\subsection{{Gaia catalogues in $\omega$~Cen}}\label{sec:fpr}
{In the Gaia DR3 source catalog there are no counterparts to the binary sources listed in Table \ref{tab:info} when we impose a maximum separation of 5\arcsec, a Gaia magnitude $m_G<20$ and magnitude difference $|m_V-m_G|<2$, and the presence of a Gaia parallax measurement.}

{On 10 October 2023, however, \gaia published a Focused Product Release (FPR) that includes an catalogue of additional sources in $\omega$~Cen \citep{2023arXiv231006551G}. In that catalogue (\gaia archive table \texttt{gaiafpr.crowded\_field\_source}, epoch J2017.5), we found unique positional matches to all five sources within a radius of 30 mas. We did not correct for proper motion between the DR1 and FPR epochs. The details of these matched sources are given in Table \ref{tab:fpr}. They have small magnitude differences $|m_V-m_G|<0.3$\,mag and the next closest positional matches are at separations $\gtrsim 0.3\arcsec$. We briefly explore the properties of these \gaia data here.}
 
\begin{table}[h!]
\footnotesize 
\centering 
  \caption{{\gaia FPR identifiers and relevant properties. The $\rho$ column indicates the match distance and the AEN column lists the FPR `astrometric\_excess\_noise`.}}
  \label{tab:fpr}
\begin{tabular}{rrrrrr}
\hline
\hline
ID & Gaia FPR source\_id & $m_V$ & $m_G$ & $\rho$ & AEN \\
  &  & (mag) & (mag) & (\arcsec) & (mas) \\
\hline
143023 & 6083701663102358400 & 17.056 & 17.023 & 0.026 & 2.02 \\
212028 & 6083701972347609216 & 18.073 & 17.869 & 0.013 & 20.90 \\
233697 & 6083701456942300288 & 17.398 & 17.363 & 0.019 & 2.92 \\
290133 & 6083701800545532544 & 18.148 & 17.913 & 0.016 & 15.97 \\
317591 & 6083698948678055168 & 18.811 & 18.820 & 0.022 & 5.36 \\
\hline
\end{tabular}
\end{table}

\begin{figure}[ht!]
\includegraphics[trim={0 0 0cm 0cm},clip, width=\columnwidth]{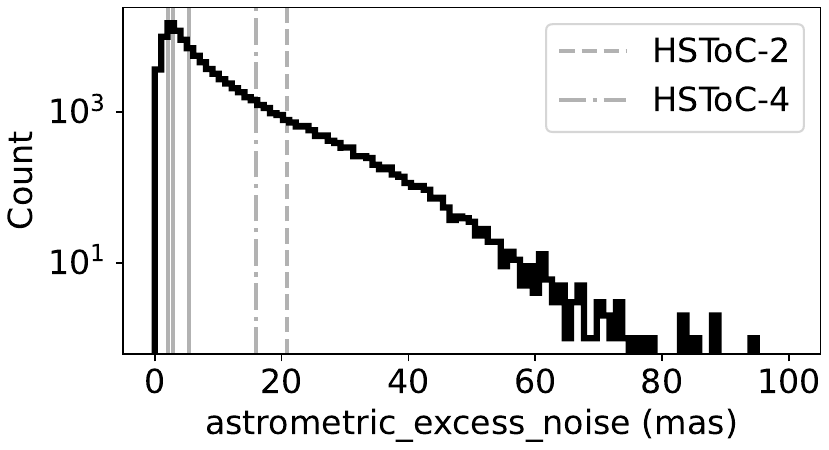}
\caption{{Histogram of the `astrometric\_excess\_noise` for 98,179 FPR sources with $17<m_G<19$, `astrometric\_params\_solved` = 95, and `astrometric\_n\_good\_obs\_al` $>$ 50. The vertical grey lines indicate the values for our five binaries.}}
\label{fig:fpr}
\end{figure}

{Figure \ref{fig:fpr} shows the FPR `astrometric\_excess\_noise` (AEN) parameter for our binaries in comparison with roughly comparable sources in that catalogue. This parameter corresponds to a jitter term that is added quadratically to the individual Gaia measurement uncertainties to account for the residual scatter in the single-source astrometric solution \citep{Lindegren:2021vn}. Without access to the Gaia astrometric timeseries it is a powerful metric to identify and sometimes even characterise astrometric binaries \citep[e.g.][]{2019A&A...631A.125K,2020MNRAS.496.1922B,2022MNRAS.510.3885G}.}

{Whereas \onefour/143023, \twothree/233697, and \threeone/317591 exhibit values close to the mode of the distribution, \twoone/212028 and \twonine/290133 show elevated levels of Gaia excess noise. These latter sources are also the ones with the larger orbital signatures ($a_1\simeq0.9$\,mas) inferred from the HST data compared to \twothree and \threeone with $a_1\simeq0.5$\,mas. This is a strong indication that the Gaia FPR is sensitive to orbital motion with amplitudes of $\sim$1 mas in $\omega$ Cen.} 

{The foreground source \onefour does, however, not match that pattern since we estimated an orbit size of $\sim$0.9 mas, but its AEN is unremarkable. This may be due to the long (and uncertain) period of this binary compared to the FPR timespan of 5 years, which reduces the probability that orbital acceleration shows up as elevated AEN.}

{It is evident in Figure \ref{fig:fpr} that there are many sources with AEN values similar or even much larger than \twothree and \threeone. A fraction of those are likely of instrumental origin, e.g.\ source blending, but another fraction could indicate acceleration signals. Using the Gaia FPR astrometric excess noise may therefore be a promising avenue for the discovery of new astrometric binaries in $\omega$ Cen, especially in regions not covered by the \hst observations.}

\section{MUSE observations}
The sources 212028 and 290133 are covered by the footprint of the
mosaic observed with MUSE \citep{muse} as part of the survey of
Galactic globular clusters presented in \citet{kamann2018}. The
quality cuts routinely performed prior to searches for binary
stars \citep[see][]{giesers2019} result in sets of six radial velocity
(RV) measurements for 212028 and five RV measurements for 290133,
with typical uncertainties of $>10\,{\rm km\,s^{-1}}$ for the former
and $5-10\,{\rm km\,s^{-1}}$ for the latter. None of the stars shows
signs of variability in the MUSE data. However, given the relative
large uncertainties compared to the expected velocity semi-amplitudes (Table \ref{tab:orbit_parameters}),
this does not seem unexpected.

We note that the spatial resolution of the MUSE data, with a typical
FWHM of $\sim0.8\,{\rm arcsec}$, is lower compared to the HST data by a
factor of $\sim$8. Therefore, even though the stars are bright enough to
perform accurate RV measurements with MUSE, crowding limits our
abilities to do so. Both sources have similarly bright stars located at
about $0.5\times$ the FWHM and significantly brighter neighbours at
distances of $\sim1\arcsec$. Future observations with the MUSE narrow
field mode will be able to overcome this problem.

\section{Discussion and future studies}\label{sec:Discuss}
\begin{figure}[h!]
\includegraphics[trim={2cm 5cm 2cm 2cm},clip, width=\columnwidth]{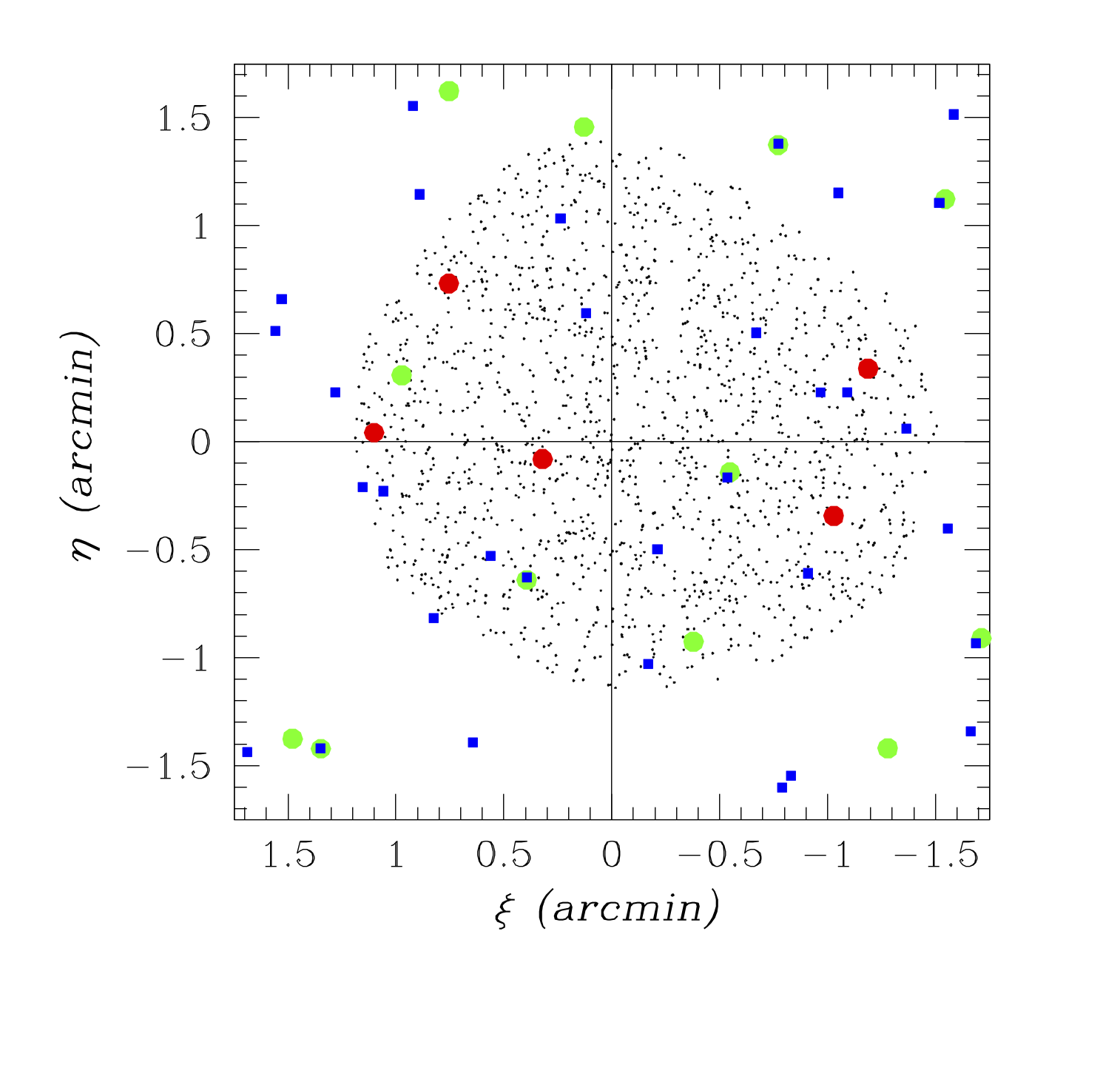}
\caption{Distribution of rare objects in $\omega$~Cen. The center of the gnomonic projection is the best-known kinematic center of $\omega$~Cen. Black dots show randomly-chosen 7\% of all program stars. Red points are oue newly-discovered binaries. The one in the upper-right quadrant is the foreground field binary. Green points indicate millisecond pulsars and blue ones X-ray sources.}
\label{fig:rare_objects}
\end{figure}

The largest Milky Way globular cluster $\omega$~Cen now is the first cluster with detected astrometric accelerations. Whether or not the relatively small range of the deduced masses of the invisible components is typical for Milky Way globular clusters remains to be answered by future studies. However, we can get some cues from another rare populations such as millisecond pulsars (MSP) and X-ray sources. Curiously, the first bona fide  MSPs in $\omega$~Cen were discovered only in 2020 \citep{dai20} followed by observations with the MeerKAT radio telescope and discovery of 13 more MSPs \citep{che223}. Four of them are in our field (Figure~\ref{fig:rare_objects}). 

Since MSPs and neutron stars share the same physical phenomenon observed in different wavelengths, we inspected the known binary systems anywhere in the sky with well-measured component masses \citep[Table~1][]{oze16}. These binary systems yield fairly-tight mass limits of $1.38\pm0.05 M\odot$. Two of our targets, 212028 and 290133, fit very well to these mass limits and thus provide a strong argument that their invisible component could be
a neutron star. In fact, the estimated properties of these two systems are very similar in luminosity, acceleration, component masses, and Keplerian parameters such as the semi-major axis.

An alternative option for the nature of these companions would be a tight pair of white dwarfs, but such a scenario may not be effective in the cores of globular clusters. The presence of MSPs in our field itself is auspicious to the neutron star scenario.

Old stellar systems such as globular clusters contain low-luminosity X-ray sources. The most
comprehensive rework of archival $\omega$~Cen observations with the {\it Chandra X-ray Observatory} Advanced CCD Imaging Spectrometer provide the best sample of these sources \citep{hen18}.
There are 14 X-ray sources within the boundaries of our field. None of them is close to the detected astrometric binaries. In globular clusters the typical X-ray sources are cataclysmic variables \citep{bel21}. They are relatively tight binaries composed of a white dwarf or a neutron star and a main sequence star as donor. Our field has four cataclysmic variables but their potential semi-major axis would be totally imperceptible.

{Future studies can expand on our work in several ways. Including archival \hst ACS/WFC images of $\omega$~Cen taken prior to WFC3/UVIS images and accumulating more data from the ongoing WFC3/UVIS calibration program will expand the time baseline of the observations and therefore lead to better constraints on the binaries' orbital and mass parameters.}

{Independently, one could experiment with relaxing the conservative cutoff value for the image quality parameter \texttt{qfit} to increase the number of available measurements per source, hence potentially increase sensitivity.} 

{Access to the epoch positions $X_f$,$Y_f$, free of the effect of proper motions using \citet{bel17} data, and new calibrated magnitudes will be granted upon request to the authors and made available through the SciServer platform.}

{Further exploration of the \gaia FPR dataset of $\omega$~Cen has the potential to lead to the discovery of additional binaries in this cluster.} 

\section{Summary and conclusions}\label{sec:conclusions}
This study is trailblazing astrometric acceleration searches in a relatively distant globular cluster populated by roughly solar- and subsolar-mass sources. We demonstrated the viability of this project by obtaining high-precision astrometric timeseries for tens of thousands of sources, thereby improving the proper motions of more than 22,000 cluster members with an average precision of 0.011 mas~yr$^{-1}$. We discovered five new binaries, one in the foreground and four cluster-members, and characterised their Keplerian motion.

Despite the large distance of $\omega$~Cen of over 5~kpc we were able to detect the first astrometric binaries. Those have very-small orbital semi-major axes below one 1~mas (Table~\ref{tab:orbit_parameters}). These binaries have periods longer than our survey timespan, except for {\twothree}, where we observed more than one Keplerian orbit allowing us to derive better constraints on the mass of the companion. {Inspection of the \gaia FPR dataset of $\omega$~Cen independently supports the discovery of the larger-amplitude binaries \twoone and \twonine, which have elevated levels of \gaia astrometric excess noise.}
 
 Our preliminary best-fit mass estimates for the invisible companions in $\omega$~Cen range from $\sim$0.7$M_{\sun}$ to $\sim$1.4$M_{\sun}$  and from $\sim$0.6$M_{\sun}$ to $\sim$2.1$M_{\sun}$ if we account for the associated 1-$\sigma$ uncertainties. The true range of allowed companion masses is likely even larger, but there are essentially three likely scenarios of their nature: a single white dwarf\footnote{According to \citet{bel17} the top of white dwarf sequence in  $\omega$~Cen is at $m_{F606W}$ $\sim$22 mag which is $\sim$4 mag fainter than the faint limit of our binaries.}, a pair of white dwarfs, or a neutron star. Because of the uncertainties in our estimates and the Chandrasekhar limit on the maximum mass of white dwarfs at 1.4 $M_\sun$, we are unable to pinpoint the most-likely scenario. 
  
Our initial goal was to probe the existence of stellar-mass black holes in binaries of $\omega$~Cen.  Although we have not unambiguously identified a BH candidate, we have demonstrated that surveys like ours are sufficiently sensitive to discover them if they exist. A system akin to Gaia BH2 at the distance of $\omega$~Cen generates an astrometric signal with a semi-amplitude of $\sim$0.9 mas and a period of 3.5 years, which lies well within the sensitivity limit of our survey, cf.\ Table \ref{tab:orbit_parameters}. Shorter period systems akin to Gaia BH1 would generate a signal of $\sim$0.25 mas and a period of 0.5 years, thus are more challenging to detect.

\section{Acknowledgments}
\begin{acknowledgments}
I.P.\ thanks Andrea Bellini for sharing the astro-photometric catalog, Timothy Brandt for testing a target, Terrence Girard for the parallax-related code and general advising. J.S.\ thanks Jari Kajava for checking for X-ray and radio counterparts to our binaries.

This research makes use of the SciServer platform. The SciServer is a collaborative  research environment for large-scale data-driven science. It is being developed at, and administered by, the Institute for Data Intensive Engineering and Science at Johns Hopkins University.
The authors gratefully acknowledge grant support for \hst program AR-16629, provided by NASA through grants from the Space Telescope Science Institute, which is operated  by the Association of Universities for Research in Astronomy, Inc., under NASA contract NAS~5-26555. 

The \hst data underlying this research are available at MAST: \dataset[10.17909/dwtd-kr81]{\doi{10.17909/dwtd-kr81}}.

This research made use of pystrometry, an open source Python package for astrometry timeseries analysis \citep{johannes_sahlmann_2019_3515526}.

This work has made use of data from the European Space Agency (ESA) mission \gaia (\url{https://www.cosmos.esa.int/gaia}), processed by the \gaia Data Processing and Analysis Consortium (DPAC, \url{https://www.cosmos.esa.int/web/gaia/dpac/consortium}). Funding for the DPAC has been provided by national institutions, in particular the institutions participating in the \gaia Multilateral Agreement.
\end{acknowledgments}

\vspace{3mm}
\facilities{Hubble Space Telescope (WFC3/UVIS), Gaia}

\appendix

\section{The foreground binary star \onefour}\label{sec:foreground} 
\begin{figure}[ht!]
\plotone{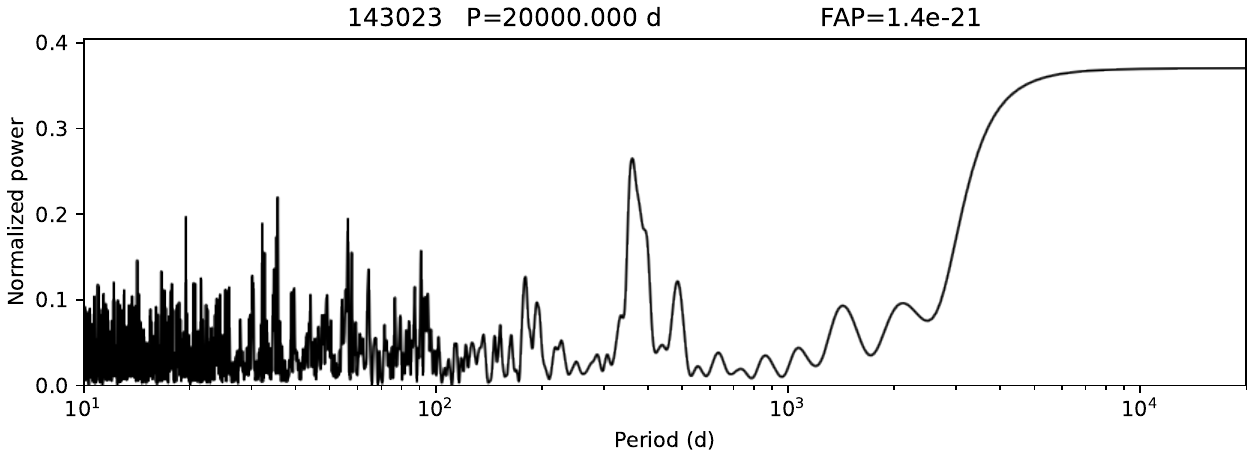}
\plotone{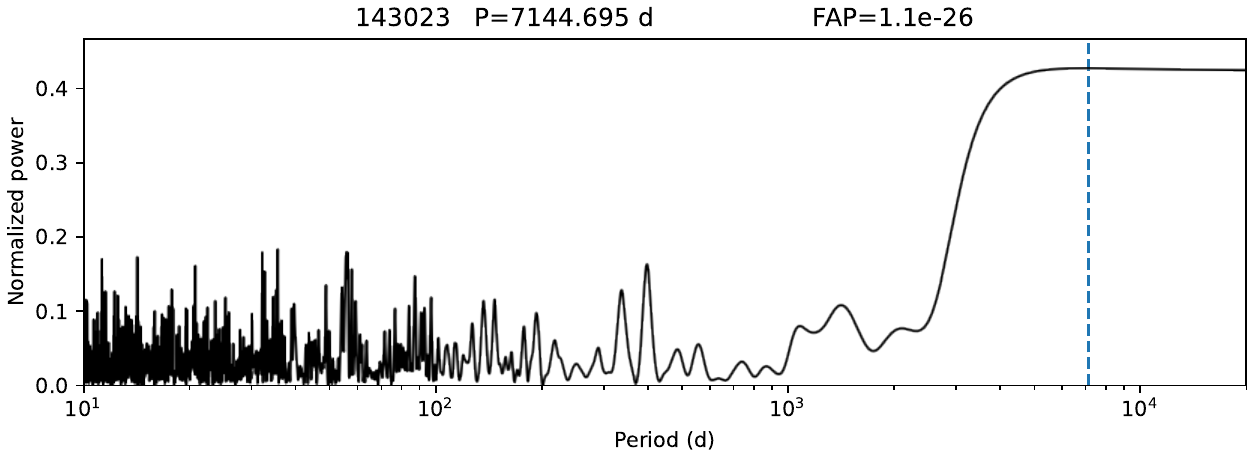}
\caption{Residual periodogram of the four- (top) and five-parameter (bottom) models for \onefour/143023. The power at long period is hardly affected but including the parallax in the model makes the power around 365 days disappear.} 
\label{fig:143023_periodograms}
\end{figure}

The astrometry periodogram of source \onefour/143023 after fitting the four-parameter linear model reveals a strong signature at a one-year period which persists in the residuals of tentative orbit-model fits. This is a strong indication that this is a foreground source whose un-modelled parallax motion introduces the periodicity. We therefore computed the parallax factors using the formalism provided by \citet{van81}  and then fitted the standard five-parameter model. As shown in Fig. \ref{fig:143023_periodograms} this cancels the power at one year. We therefore modelled the astrometry of this source with models that include the parallax term.

To estimate the potential orbital parameters, we modelled the data with the combination of the standard five-parameter model and a single Keplerian model with initial parameters compatible with the highest periodogram-peak. Since the orbital motion is only partially covered (Fig. \ref{fig:4_accl}) we can only determine a lower limit to the period and give indicative bounds to the possible mass range of the companion. Since orbital eccentricity is essentially unconstrained at this point, we limited its allow range to $<$0.3 (see Table~\ref{tab:orbit_parameters}). 

\begin{figure}[ht!]
\includegraphics[trim={0 0 15cm 0.6cm},clip, width=0.5\columnwidth]{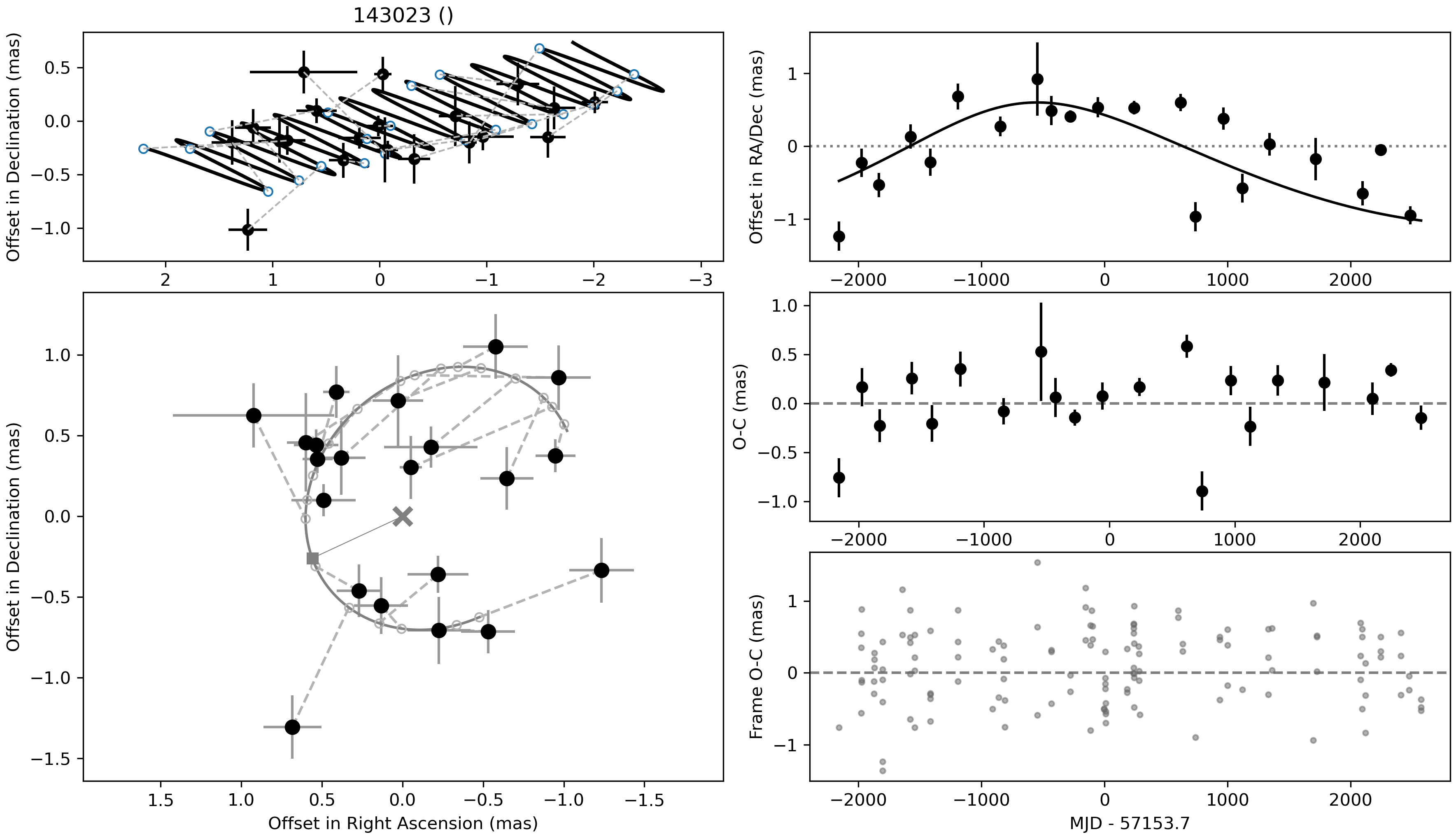}
\caption{Parallax and proper motion (top) and orbital motion (bottom) of \onefour in the sky. } 
\label{fig:143023_orbit_sky}
\end{figure}

\begin{figure}[ht!]
\centering
\includegraphics[trim={0 0 0cm 0.5cm},clip, width=0.5\columnwidth]{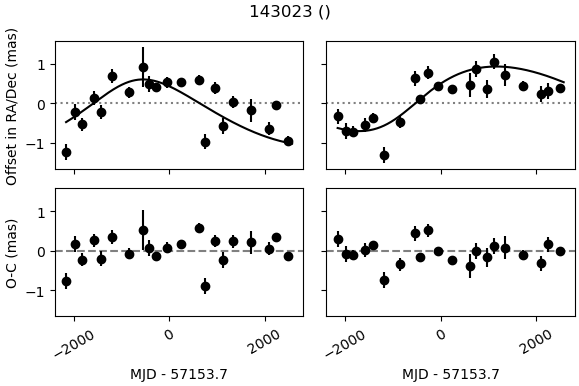}
\caption{Orbital motion of \onefour 
after subtracting the standard model (top panels). The standard deviation of 0.59 mas (both axes, computed on normal points) in the top panels is is reduced to 0.32 mas when including the Keplerian model.}
\label{fig:143023_orbit_time}
\end{figure}

One possible orbital solution with a period of $\sim7000$ days is shown in Fig. \ref{fig:143023_orbit_sky}. However, the orbital period could be as short as the $\sim4000$ day timespan of the data or much longer. The orbital motion and fit residuals for this solution as a function of time are shown in Figure \ref{fig:143023_orbit_time}.

The determined parallax of the system is almost independent of the orbital parameters because the timescales are very different for such long-period systems. We determined a relative parallax of $\varpi_\mathrm{rel}=0.55 \pm 0.12$ mas. Since that is measured relative to cluster members, we need to account for the cluster distance of 5.24 kpc \citep{sol21} and obtain an absolute parallax of \onefour of $\varpi_\mathrm{abs}=0.73 \pm 0.12$ mas.

Adopting this latter parallax value, and assuming that the \onefour primary is a metal-poor halo star with the same foreground extinction as $\omega$~Cen, its placement in theoretical isochrones provides a mass estimate in the range between 0.65 and 0.71 $M_\sun$.
Assuming a primary mass of 0.68 , the companion mass for this particular solution would be $\sim$0.15$M_{\sun}$, where we neglect any light contribution by the companion, which could dilute the photocentre motion and lead to an underestimation of the companion mass (Table~\ref{tab:orbit_parameters}).

\section{Additional figures}
\begin{figure}[ht!]
\centering
\includegraphics[trim={0 0 0cm 0.5cm},clip, width=0.5\columnwidth]{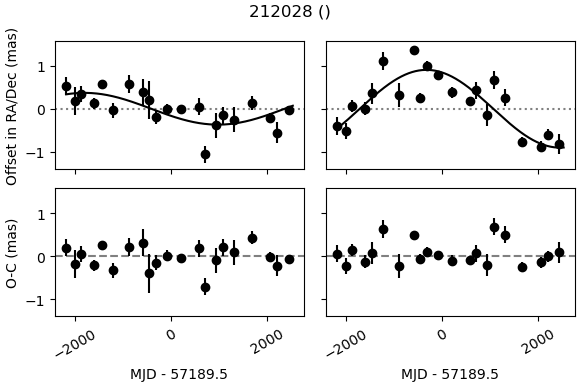}
\caption{Orbital motion of \twoone 
after subtracting the standard model (top panels). The standard deviation of 0.38 mas (both axes, computed on normal points) in the top panels is is reduced to 0.27 mas when including the Keplerian model.} 
\label{fig:212028_orbit_time}
\end{figure}

\begin{figure}[ht!]
\centering
\includegraphics[trim={0 0 0cm 0.5cm},clip, width=0.5\columnwidth]{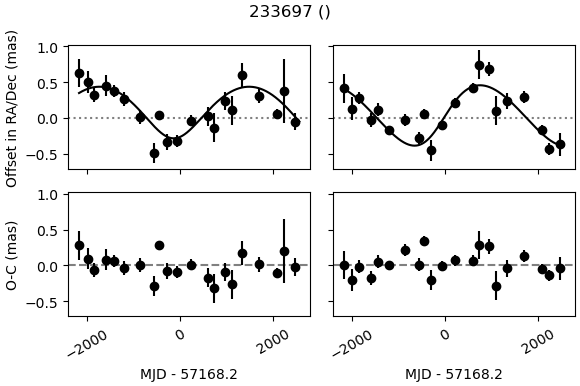}
\caption{Orbital motion of \twothree 
after subtracting the standard model (top panels). The standard deviation of 0.30 mas (both axes, computed on normal points) in the top panels is is reduced to 0.16 mas when including the Keplerian model.} 
\label{fig:233697_orbit_time}
\end{figure}

\begin{figure}[ht!]
\centering
\includegraphics[trim={0 0 0cm 0.5cm},clip, width=0.5\columnwidth]{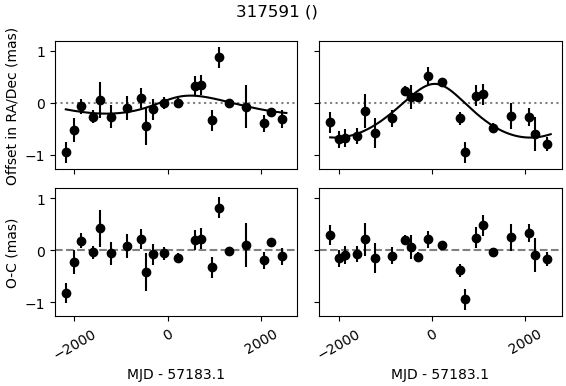}
\caption{Orbital motion of \threeone 
after subtracting the standard model (top panels). The standard deviation of 0.36 mas (both axes, computed on normal points) in the top panels is is reduced to 0.31 mas when including the Keplerian model.} 
\label{fig:317591_orbit_time}
\end{figure}

\bibliography{biblio}{}
\bibliographystyle{aasjournal}

\end{document}